\documentclass{article}

\pdfoutput=1
\usepackage{lmodern}
\usepackage{float}
\usepackage{microtype}
\usepackage{xspace}
\usepackage{amsmath,amssymb,amsfonts,amsthm,mathtools}
\usepackage{mathrsfs,bm}
\usepackage[colorlinks=true, linkcolor = blue, hyperfootnotes=false, citecolor = blue]{hyperref}
\usepackage{natbib}
\usepackage[T1]{fontenc}
\usepackage[latin1]{inputenc}
\usepackage[english]{babel}
\usepackage{graphicx}
\renewcommand{\[}{\begin{equation}}
\renewcommand{\]}{\end{equation}}
\newcommand{\abs}[1]{\left\lvert #1 \right\rvert}
\newcommand{\scalar}[1]{\left\langle{#1}\right\rangle}
\newcommand{\tensorial}{\!\otimes\!}
\renewcommand{\t}[1]{{\bm{#1}}}
\newcommand{\curve}{\mathscr{C}}
\renewcommand*{\Im}{\operatorname{Im}} 
\renewcommand*{\Re}{\operatorname{Re}} 
\renewcommand{\refeq}[1]{(\ref{#1})}
\newcommand{\dd}{\,\mathrm{d}}

\begin{document}
\title{Topological mechanics of edge waves in Kagome lattices}

\author{H. Chen, H. Nassar\footnote{Corresponding authors: HN (nassarh@missouri.edu), GH (huangg@missouri.edu)}, G.L. Huang$^*$}
\date{\small
Department of Mechanical and Aerospace Engineering, University of Missouri, Columbia, MO 65211, USA}

\maketitle
\begin{abstract}
Topological insulators are new phases of matter whose properties are derived from a number of qualitative yet robust topological invariants rather than specific geometric features or constitutive parameters. Here, Kagome lattices are classified based on a topological invariant directly related to the handedness of a couple of elliptically polarized stationary eigenmodes in the context of what is known as the ``quantum valley Hall effect'' in physics literature. An interface separating two topologically distinct lattices, i.e., two lattices with different topological invariants, is then proven to host two topological Stoneley waves whose frequencies, shapes and decay and propagation velocities are quantified. Conversely, an interface separating two topologically equivalent lattices will host no Stoneley waves. Analysis is based on an asymptotic model derived through a modified high-frequency homogenization procedure. This case study constitutes the first implementation of the quantum valley Hall effect in in-plane elasticity. A preliminary discussion of 1D lattices is included to provide relevant background on topological effects in a simple analytical framework.
\end{abstract}

\section{Introduction}
The coupling that occurs at the free boundary of an elastic solid between pressure (P) and shear vertical (SV) waves famously gives rise to a class of surface waves first discovered by Lord \citet{Rayleigh1885}. Similar mechanisms are at the origin of the interface waves known as \citet{Stoneley1924} waves and propagated along a discontinuity surface separating two distinct elastic solids. Recently, with the advent of electronic topological insulators and of their mechanical counterparts, a novel family of surface and interface waves, qualified as ``topological'', has emerged \citep{Hasan2010,Qi2011,Huber2016}. In contrast to their predecessors, topological Rayleigh and Stoneley waves ($i$) are characterized through robust topological quantities, referred to as ``invariants'', rather than by algebraic dispersion relations; ($ii$) exist in frequency bands that are bandgaps for the underlying half-space(s); and ($iii$) are immune to backscattering by a class of defects. This last feature in particular turns the edge of a topological insulator practically into a one-way waveguide with superior transmission qualities.

Following their discovery in electron lattices, topological insulators rapidly spread to photonic, acoustic and phononic crystals and metamaterials by adapting relevant quantum mechanical tools and concepts such as geometric phases, Chern numbers and the adiabatic theorem (see, e.g., \citealp{Berry1984}, \citealp{Xiao2010} and \citealp{Nassar2018}). Although remarkably fruitful, the use of quantum mechanical vocabulary can hinder the expansion of genuine solid mechanical approaches. Elaborating such an approach is the main purpose of the paper. It is presented here as a case study of in-plane topological Stoneley waves in 2D Kagome lattices in the context of what is known as valleytronics or ``quantum valley Hall insulator'' in physics literature \citep{Neto2009}. Similar insulators have been previously investigated for acoustic waves \citep{Lu2016a,Lu2016,Ni2017} and out-of-plane flexural waves \citep{Vila2017,Pal2017,Liu2017} whereas other insulators, such as the quantum Hall \citep{Yang2015,Nash2015,Fleury2016,Chen2016c} and the quantum spin Hall insulators \citep{Susstrunk2015,Mousavi2015,Yves2017,Yves2017a}, have been implemented in more general 2D geometries.

It is common to explore free wave propagation in the bulk of phononic crystals and metamaterials through their dispersion diagrams $\omega=\omega(q)$. This allows in particular to determine phase and group velocities as well as the location and width of potential bandgaps. In comparison, the study of bulk eigenmodes $\Psi=\Psi(q)$ as a function of wavenumber $q$ had no foreseeable consequences. It was the main contribution of topological methods to show that the way in which the phase profile of $\Psi(q)$ changes as $q$ goes through the Brillouin zone is deeply connected to the existence of edge states within bandgaps. In the present paper, this connection, known as the principle of bulk-edge correspondence, is exemplified through a careful scrutiny of the eigenmode shapes and their corresponding orbits.

More specifically, Kagome lattices will be classified based on a topological invariant directly related to the sign of a physical contrast parameter and to the handedness of a couple of elliptically polarized stationary eigenmodes. An interface separating two topologically distinct lattices, i.e., two lattices with different topological invariants, is then proven to host two Stoneley waves whose frequencies, shapes and decay and propagation velocities are quantified. Conversely, an interface separating two topologically similar lattices will host no Stoneley waves. Calculations are based on an asymptotic model derived through a modified high-frequency homogenization procedure; see \cite{Harutyunyan2016} for a recent review of these methods. Last, immunity to backscattering by interface corners is verified thanks to numerical transient simulations. First however, a study of a 2-periodic 1D lattice introducing a number of basic concepts of topological mechanics is presented.
\section{An introduction to topological effects in 1D lattices}
1D spring-mass lattices are investigated in the context of topological mechanics. Various known results are reinterpreted using topological tools. The purpose is to demonstrate and provide a clear understanding of a number of topology concepts within a simple framework before tackling the more involved study presented in the next section.
\subsection{Governing equations}
\begin{figure}[ht!]
\centering
\includegraphics[keepaspectratio,width=\textwidth]{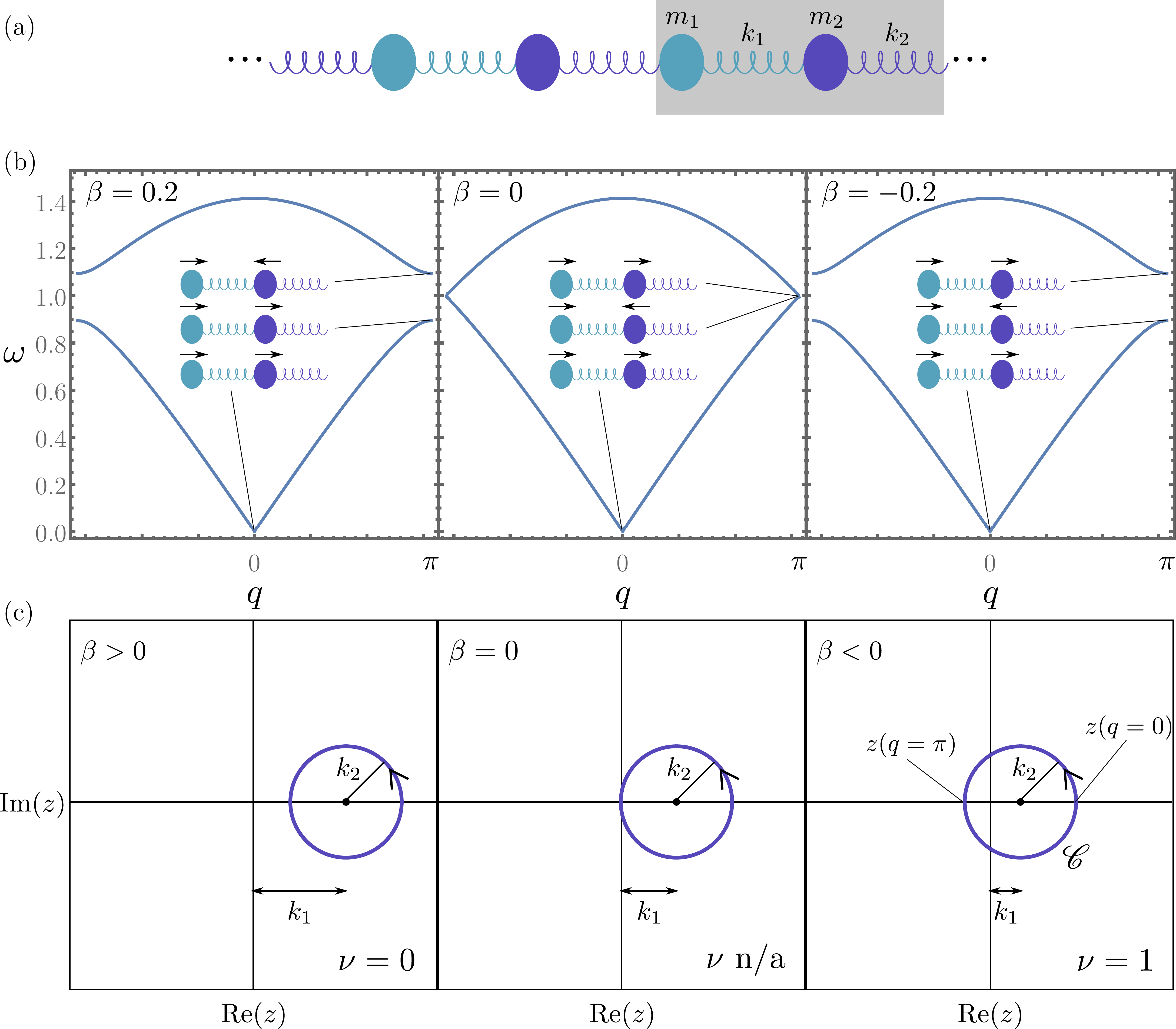}
\caption{(a) a 2-periodic spring-mass lattice: a unit cell is highlighted; (b) typical dispersion diagrams for different contrast $\beta$ and $m_1=m_2=m$: angular frequency is measured in units of $\sqrt{(k_1+k_2)/m}$. The insets illustrate the phase profile of the Floquet-Bloch eigenmodes at the pinned points. (c) The closed loop $\curve$ described by $z(q)$ as $q$ spans the Brillouin zone: $\curve$ winds once around the origin for $\beta<0$.}
\label{fig:fig1}
\end{figure}

Consider the 2-periodic spring-mass lattice of Figure~\ref{fig:fig1}a. A unit cell contains two masses $m_1$ and $m_2$ of equal value $m$ and two springs of constants $k_1$ and $k_2$. Newton's second law of motion applied for each of the masses can be expressed as
\[
\label{eq:ME}
\begin{split}
m\ddot u ^n_1 &= -(k_1+k_2)u^n_1 + k_1u_2^n + k_2u_2^{n-1},\\
m\ddot u ^n_2 &= -(k_1+k_2)u^n_2 + k_1u_1^n + k_2u_1^{n+1},
\end{split}
\]
where $u_j^n\equiv u_j^n(t)$ is the displacement of mass number $j=1,2$ of the $n^\text{th}$ unit cell at time $t$ and a superimposed dot denotes a time derivative. This infinite set of difference equations can be transformed into a $2\times 2$ eigenvalue problem using a Floquet-Bloch expansion. Thus, letting
\[
u_j^n = a_j e^{i(nq-\omega t)} \equiv a_j Q^n e^{-i\omega t}
\]
reduces the equations of motion to the matrix form
\[
-\hat H
\begin{bmatrix}
a_1 \\ a_2
\end{bmatrix}
=
-\omega^2
m
\begin{bmatrix}
a_1 \\ a_2
\end{bmatrix},
\quad
\hat H =
\begin{bmatrix}
k_1+k_2 & -k_1-k_2Q^* \\ -k_1-k_2Q & k_1 + k_2
\end{bmatrix},
\]
where $\omega$ is angular frequency, $q$ is dimensionless wavenumber and $Q=e^{iq}$ is a phase factor. Typical dispersion diagrams deduced from the dispersion relation
\[
\det\left(\hat H -\omega^2 m I\right) = 0, \quad
I = \begin{bmatrix}
1 & 0 \\ 0 & 1
\end{bmatrix},
\]
are depicted in Figure~\ref{fig:fig1}b for varying contrast $\beta \equiv (k_1-k_2)/(k_1+k_2)$ and constant offset $(k_1+k_2)/2$. These are reinterpreted next using the language of topology.
\subsection{Band inversion and topology}
Inspecting the dispersion diagrams, configurations $\beta=\beta_0\neq 0$ and $\beta=-\beta_0$ seem identical but are in fact distinct. The difference lies in the shape of the eigenmodes. For instance, following the acoustic branch, the masses within one unit cell oscillate in phase for $q\approx 0$ and remain oscillating in phase as $q$ approaches $\pi$ at the boundary of the Brillouin zone if $k_1>k_2$, or $\beta>0$ (Figure~\ref{fig:fig1}b). When in contrast $k_1<k_2$, or $\beta<0$, the masses are perfectly in phase at $q=0$ and are in opposition of phase at $q=\pi$. This is because the eigenmodes of the less energetic acoustic branch attempt to localize deformations in the softest of the two springs, $k_2$ in the first case and $k_1$ in the second case, while leaving the stiffest one undeformed.

Therefore, as $\beta$ changes from positive to negative values, the acoustic eigenmode at $q=\pi$ ``twists'' and changes its shape from $(a_1=1/\sqrt{2},a_2=1/\sqrt{2})$ to $(a_1=1/\sqrt{2},a_2=-1/\sqrt{2})$; see Figure~\ref{fig:fig1}b. The transition occurs exactly at $\beta=0$ when the gap closes. This ``twisting'' phenomenon is referred to as a ``band inversion''. The fact that band inversion is accompanied here by the gap closing might seem accidental for there is no apparent reason why changing the phase of an acoustic mode should require the gap to close. Interestingly, topological considerations show that, indeed, the described band inversion phenomenon cannot be completed \emph{without closing the gap} regardless of how $k_1$ and $k_2$ are changed. This is proven next.

Let us quantify the change in phase $\Delta$ that the acoustic eigenmode incurs in going from $q=0$ to $q=\pi$. Recall that the phase difference between two unitary complex numbers $z_1=e^{i\alpha}$ and $z_2=e^{i\beta}$ can be obtained as
\[
\beta-\alpha \approx \sin(\beta-\alpha) = \Im(z_1^*z_2) = \Im(z_1^*\delta z)
\]
when $\delta z =z_2-z_1 \approx 0$. Similarly, one has
\[
\Delta = \Im\int_0^\pi \scalar{\Psi,\partial_q\Psi}\dd q
\]
where $\scalar{}$ is the usual Hermitian dot product, $\Psi\equiv\Psi(q)$ is the normalized acoustic eigenmode at wavenumber $q$ and the derivative $\partial_q\Psi$ quantifies the change in $\Psi$ due to a change in $q$ in going from $0$ to $\pi$. Letting $z(q)=k_1 + k_2e^{iq}$ be the off-diagonal term in the dynamical matrix $\hat H$, one has
\[
\Psi =
\frac{1}{\sqrt 2}
\begin{bmatrix}
1 \\
z/\!\abs{z}
\end{bmatrix}
\]
so that
\[
\Delta = \frac{1}{2}\Im\int_0^\pi \frac{z^*}{\abs{z}}\partial_q\left(\frac{z}{\abs{z}}\right)\dd q.
\]
Upon expanding the derivatives and observing that $z^*(q)=z(-q)$, the expression of $\Delta$ can be transformed into the simple form
\[
\Delta = \frac{1}{4}\Im\int_{-\pi}^\pi \frac{\partial_q z}{z}\dd q = \frac{1}{4}\Im\int_\curve \frac{\dd z}{z} = \frac{\pi \nu}{2}
\]
where $\curve$ is the oriented curve that $z(q)$ describes as $q$ spans the Brillouin zone $[-\pi,\pi]$. One recognizes above that $\Delta$ is $\pi/2$ times the winding number $\nu$ of curve $\curve$ around the origin $0$ (Figure~\ref{fig:fig1}c). Given that $z(q=-\pi)=z(q=\pi)$, curve $\curve$ is in fact a closed loop meaning that $\nu$ is necessarily an integer. One can now conclude thanks to the following line of reasoning: band inversion amounts to changing $\Delta$ from $0$ to $\pi/2$, i.e., to changing $\nu$ from $0$ to $1$. This requires somehow displacing the origin from outside loop $\curve$ to its inside which cannot be completed without $\curve$ crossing the origin at which point $z$ vanishes and the acoustic and optical branches touch. All in all, the acoustic band cannot be ``inverted'' by continuously perturbing the spring constants without closing the bandgap.

\begin{figure}[ht!]
\centering
\includegraphics[keepaspectratio,width=\textwidth]{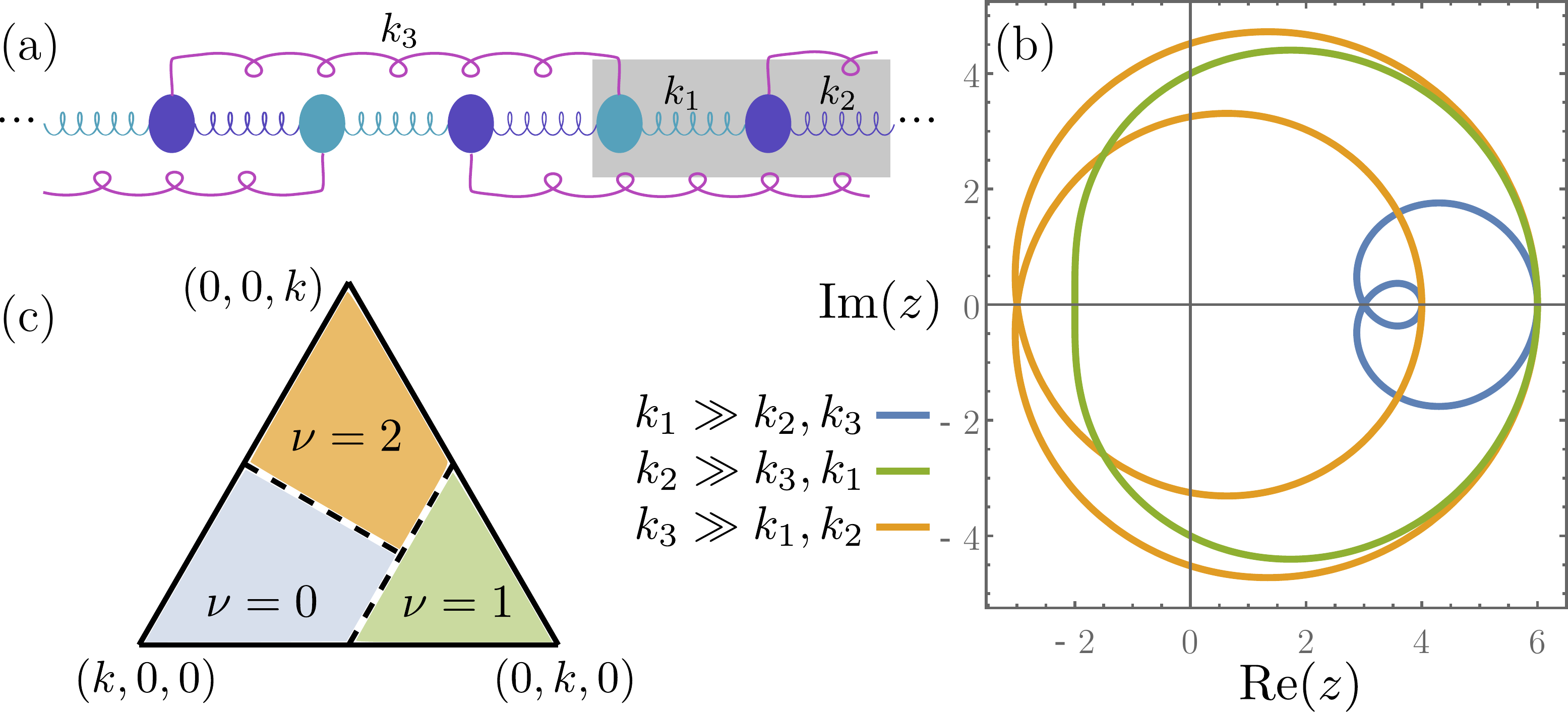}
\caption{(a) Example of a 2-periodic spring-mass lattice with next-nearest-neighbor interactions. (b) Loop $\curve$ is described by $z(q)=k_1+k_2Q+k_3Q^2$ and wraps zero, one or two times around the origin respectively when $k_1$, $k_2$ or $k_3$ is dominant. (c) Phase diagram of the winding number in the $(k_1,k_2,k_3)$-space projected onto the plane $k_1+k_2+k_3=k$; the winding number in not defined over dashed lines where the gap closes.}
\label{fig:highWind}
\end{figure}

The phase accumulated by $\Psi$ in going from $0$ to $\pi$ can grow beyond $\pi/2$ when next-nearest-neighbor interactions are allowed. In that case, the acoustic branch ``twists'' more than once and the winding number $\nu$ can take larger values as $\curve$ winds more than once around the origin. An example is provided in Figure~\ref{fig:highWind}.
\subsection{Bulk-edge correspondence}
\begin{figure}[ht!]
\centering
\includegraphics[keepaspectratio,width=\textwidth]{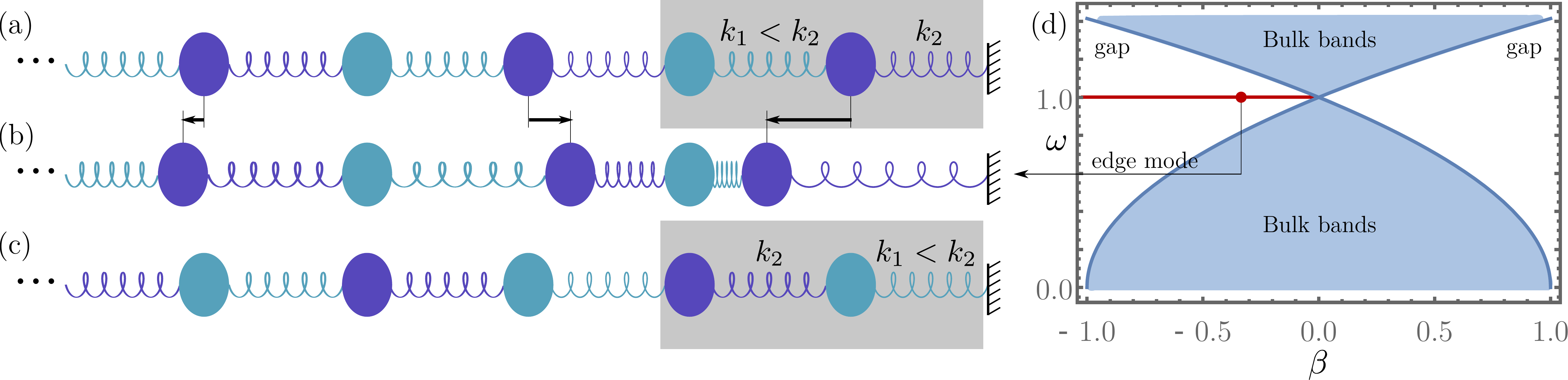}
\caption{The bulk-edge correspondence principle illustrated for a boundary: (a) a semi-infinite lattice with $\beta<0$, i.e., $\nu=1$, supports one edge mode at its fixed boundary illustrated on (b); (c) a semi-infinite lattice with $\beta>0$, i.e., $\nu=0$ supports no edge modes at its fixed boundary; (d) bulk (in blue) and edge (in red) spectra.}
\label{fig:semiInfinite}
\end{figure}

Motivated by the considerations of the previous subsection, two lattices featuring a common bandgap are called ``topologically equivalent'' if one can be continuously changed into the other without closing the gap. They are called ``topologically distinct'' if such transformation cannot be completed without closing the gap. For instance, two lattices with $k_1$ and $k_2$ swapped are topologically distinct even though one can be deduced from the other by a simple redrawing of the unit cell. This unsettling observation takes full sense when dealing with \emph{finite} lattices where in the vicinity of a given boundary, the lattice terminates in a unique fashion allowing to choose the unit cell unambiguously (Figure~\ref{fig:semiInfinite}a,c). By the same logic, it is at boundaries that topological inequivalence has the most important consequences gathered under the name of the ``principle of bulk-edge correspondence''. The principle states that a lattice with winding number $\nu$ supports at its boundary $\nu$ localized eigenmodes whose frequencies fall inside the bandgap. Further, two lattices with winding numbers $\nu_1$ and $\nu_2$ support at their interface $\abs{\nu_1-\nu_2}$ localized eigenmodes.
\begin{figure}[ht!]
\centering
\includegraphics[keepaspectratio,width=\textwidth]{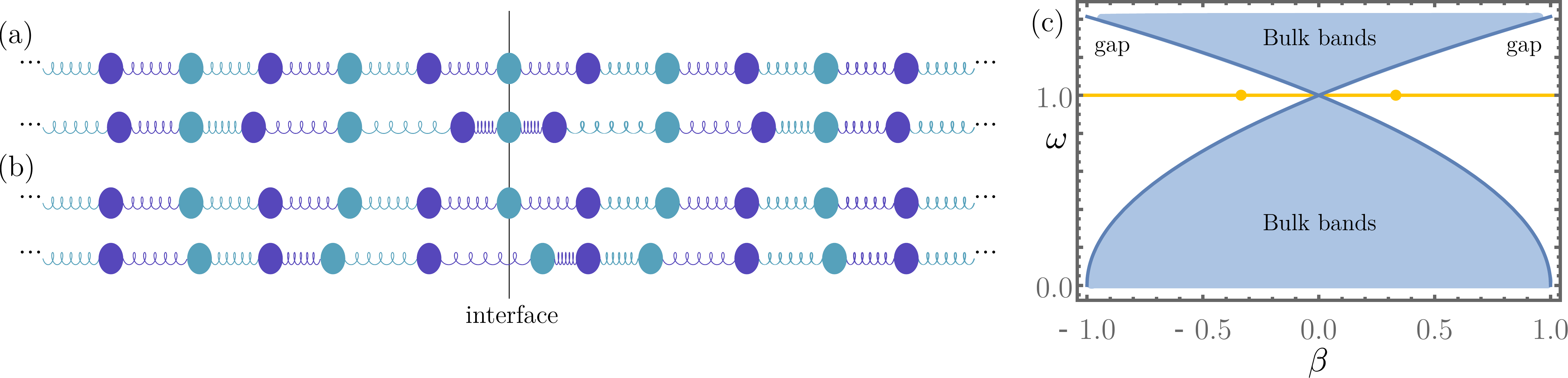}
\caption{The bulk-edge correspondence principle illustrated for an interface: (a,b) localized modes at the interface separating two topologically distinct lattices respectively for $\beta>0$ and $\beta<0$: reference state (top) and snapshot of deformed state (bottom); (c) bulk (in blue) and interface (in yellow) spectra.}
\label{fig:semiInfinite2}
\end{figure}

Consider for instance the semi-infinite lattice ($n<0$) of Figure~\ref{fig:semiInfinite}a with a Dirichlet boundary condition: $u_1^0=0$. Its bandgap is centered on the frequency $\omega=\sqrt{(k_1+k_2)/m}$. At that frequency, the motion equations~\refeq{eq:ME} uncouple and reduce to
\[
0 = k_1u_2^n + k_2u_2^{n-1},
\quad 0 = k_1u_1^n + k_2u_1^{n+1}.
\]
This readily implies that $u_1^n$ is constantly null whereas $u_2^n$ propagates following the geometric rule
\[
u_2^{n-1} = (-k_1/k_2)u_2^n.
\]
Thus, when $k_1<k_2$, $u_2^n$ decays exponentially as $n\to -\infty$ and is an admissible eigenmode localized near the boundary $n=0$ (Figure~\ref{fig:semiInfinite}b). In contrast, for $k_1>k_2$ (Figure~\ref{fig:semiInfinite}c), $u_2^n$ grows exponentially as $n\to -\infty$ and is not an admissible eigenmode. Both cases are in agreement with the bulk-edge correspondence principle as stated above. Note that for a semi-infinite lattice with $n>0$, the circumstances are inverted. Accordingly, the spectrum of a semi-infinite lattice ($n<0$) includes, in addition to the bulk acoustic and optical bands, a single frequency in the bandgap corresponding to an edge mode for $\beta<0$ (Figure~\ref{fig:semiInfinite}d).

This elementary proof of the principle generalizes rather immediately to other lattices with next-nearest-neighbor interactions and higher winding numbers. A key observation there is that the ratio $-k_1/k_2$ is a root of $z\equiv z(Q)$ understood as a polynomial in the phase factor $Q$: $z(-k_1/k_2)=0$. Then, the number of localized eigenmodes is directly related to the number of roots of $z$ that have a magnitude smaller than $1$ which is equal to the winding number $\nu$ by Cauchy's residue theorem. A complete proof will not be pursued and can be adapted from the one given by \citet{Chen2017} in a quantum mechanical context.

Now consider two semi-infinite lattices connected at mass $1$ of unit cell $n=0$. Across the interface, spring constants $k_1$ and $k_2\neq k_1$ are swapped. Thus, the left and right semi-infinite lattices are topologically distinct and have an absolute difference in winding numbers equal to $1$. One localized interface mode is therefore expected. The same motion equations as before lead to the expressions
\[
u_1^n = (-k_1/k_2)^n u_1^0, \quad
u_2^n = (-k_2/k_1)^n u_2^{\pm 1}.
\]
Of these two modes, only one survives the decay condition at infinity and corresponds to $u_1^0=0$, $u_2^{+1}=-u_2^{-1}$ if $k_1>k_2$ (Figure~\ref{fig:semiInfinite2}a) and to $u_2^{\pm 1}=0$ if $k_2>k_1$ (Figure~\ref{fig:semiInfinite2}b).
\subsection{Topological protection}
The edge modes considered above are protected against uncertainty and disorder in the values of $k_1$, $k_2$ and/or $m$: their perturbation, as long as the bandgap remains open, cannot change the winding number $\nu$ and therefore cannot change the number of supported edge modes. These edge modes can then be qualified as robust, topological or topologically protected. Note however that topological protection is not absolute and will hold as long as $\nu$ remains quantized and in particular as long as $m_1=m_2$.
\begin{figure}[ht!]
\centering
\includegraphics[keepaspectratio,width=\textwidth]{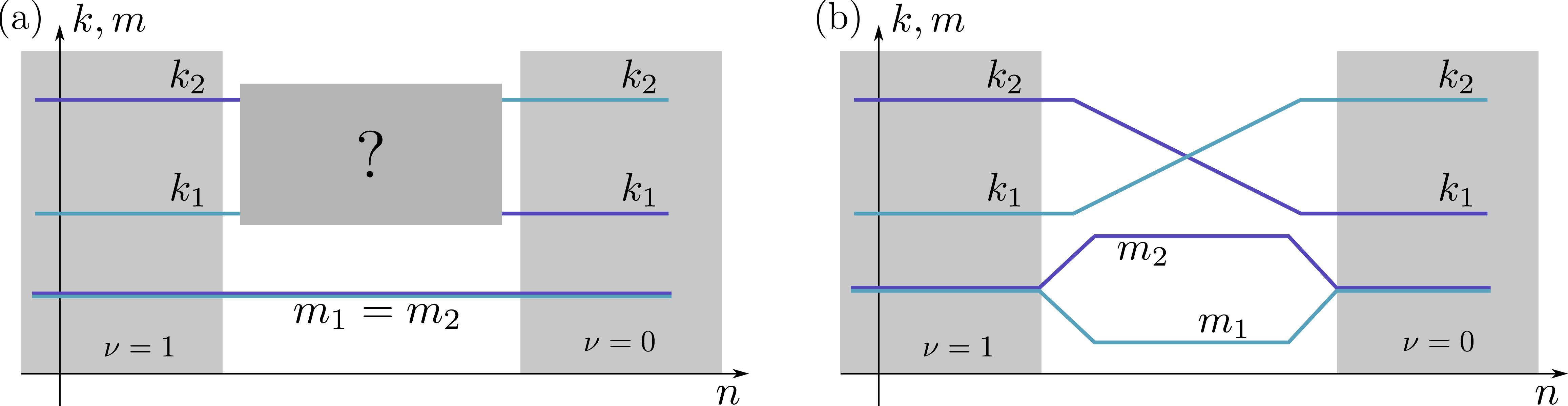}
\caption{Topological protection holds for $m_1=m_2$ (a) and fails for $m_1\neq m_2$ (b).}
\label{fig:perturbation}
\end{figure}

To illustrate that fact, consider the two scenarios illustrated on Figure~\ref{fig:perturbation}. In scenario (a), the interface will host a localized state regardless of the way in which $k_1$ and $k_2$ swap places. In scenario (b) however, as the value of $m_1$ breaks apart from $m_2$ while $k_1$ and $k_2$ cross, the width of the gap given by
\[
\delta \omega^2 = 
\frac{\abs{z}^2}{m_1m_2}+
\frac{(k_1+k_2)^2}{4}\left(\frac{1}{m_1}-\frac{1}{m_2}\right)^2
\]
remains non-zero even as the winding number changes continuously from $1$ to $0$. Thus, $\nu$ is no longer quantized and the principle of bulk-edge correspondence fails.

This observation generalizes to other kinds of topological insulators where the qualities of the system are protected by some symmetry. Here, the symmetry is $m_1=m_2$. Other examples include the quantum spin Hall effect protected by a time reversal symmetry and the quantum valley Hall effect, investigated next, protected by a $ C_3$ symmetry.
\section{Topological Stoneley waves in gapped Kagome lattices}
Interface modes in two dimensions take the form of Stoneley waves. Although classical Stoneley waves propagate at low frequencies falling within the first bulk passing band, in the following, band inversion in Kagome lattices is shown to lead to the apparition of Stoneley waves within a total bulk bandgap. In condensed matter physics literature, the phenomenon is known as ``quantum valley Hall effect'' and was investigated for acoustic and flexural waves. Next, it will be analytically and numerically demonstrated for the first time in in-plane elasticity. Derivations are based on an asymptotic homogenized model obtained first.
\subsection{Discrete model}
\begin{figure}[ht!]
\centering
\includegraphics[keepaspectratio,width=\textwidth]{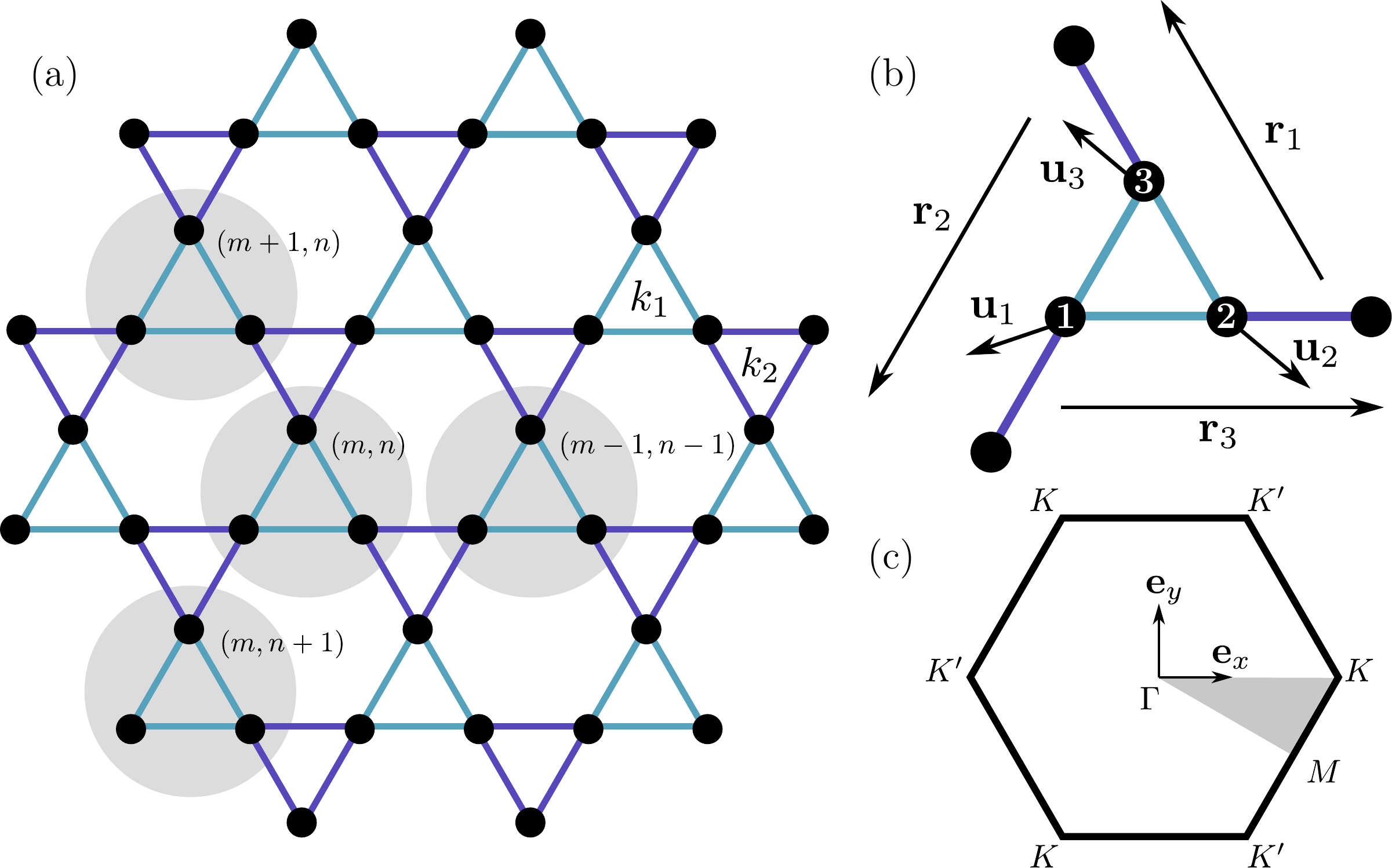}
\caption{(a) A regular Kagome lattice: edges are massless springs and nodes are perfect massive hinges; (b) an annotated unit cell; (c) the first Brillouin zone.}
\label{fig:Kagome}
\end{figure}
The Kagome lattice can be obtained by stacking copies of the 1D model investigated in the previous section in three directions $\t r_1$, $\t r_2$ and $\t r_3$ separated by an angle of $2\pi/3$. Hence,
\[
\scalar{\t r_1,\t r_2}=\scalar{\t r_2,\t r_3}=\scalar{\t r_3,\t r_1}=-1/2,\quad
\t r_1+\t r_2+\t r_3 = 0.
\]
The resulting lattice is illustrated on Figure~\ref{fig:Kagome}a. A unit cell contains three masses of value $m$ totaling six degrees of freedom and six springs of constants $k_1$ and $k_2$ (Figure~\ref{fig:Kagome}b). Letting $\t u_j^{m,n}$ be the displacement of mass $j$ in unit cell $(m,n)$, a Floquet-Bloch wave of wavenumber $\t q$ and frequency $\omega$ is characterized by
\[
\t u_j^{m,n} = \t u_j e^{i(q_1m+q_2n-\omega t)} \equiv Q_1^m Q_2^n \t u_j e^{-i\omega t}
\]
where $Q_j=e^{iq_j}$ is the phase factor gained in going one unit cell across in the direction $\t r_j$ and $q_j = \scalar{\t q,\t r_j}$, for $j=1,2,3$. Note that since $\t r_3=-\t r_1-\t r_2$, moving one cell in the direction $\t r_3$ is equivalent to moving one cell in the direction $-\t r_1$ and another in the direction $-\t r_2$. Similarly, $q_3=-q_1-q_2$ and $Q_3=Q_1^*Q_2^*$. It is then straightforward to check that the equations of motion using these notations read
\begin{gather*}
\begin{split}
-\omega^2 m \t u_1 = k_1\scalar{\t u_3-\t u_1,\t r_2}\t r_2 &+ k_2\scalar{Q_2 \t u_3-\t u_1,\t r_2}\t r_2 \\
&+ k_1\scalar{\t u_2-\t u_1,\t r_3}\t r_3 + k_2\scalar{Q_3^*\t u_2-\t u_1,\t r_3}\t r_3,
\end{split}
\\
\begin{split}
-\omega^2 m \t u_2 = k_1\scalar{\t u_1-\t u_2,\t r_3}\t r_3 &+ k_2\scalar{Q_3\t u_1-\t u_2,\t r_3}\t r_3 \\
&+ k_1\scalar{\t u_3-\t u_2,\t r_1}\t r_1 + k_2\scalar{Q_1^*\t u_3-\t u_2,\t r_1}\t r_1,\\
\end{split}
\\
\begin{split}
-\omega^2 m \t u_3 = k_1\scalar{\t u_2-\t u_3,\t r_1}\t r_1 &+ k_2\scalar{Q_1\t u_2-\t u_3,\t r_1}\t r_1\\
&+k_1\scalar{\t u_1-\t u_3,\t r_2}\t r_2 + k_2\scalar{Q_2^* \t u_1-\t u_3,\t r_2}\t r_2.
\end{split}
\end{gather*}
Recalling that $\scalar{\t u_j,\t r_k}\t r_k = \t r_{kk}\t u_j$ with $\t r_{kk}=\t r_k\tensorial \t r_k$, the motion equations can be gathered in the compact matrix form
\[\label{eq:MEK}
-\hat H \hat u = -\omega^2 m \hat u
\]
with
\[
\hat u = 
\begin{bmatrix}
\t u_1 \\ \t u_2 \\ \t u_3
\end{bmatrix}, \quad
\hat H =
\begin{bmatrix}
z_0(\t r_{22}+\t r_{33}) & -z_3^* \t r_{33} & -z_2 \t r_{22} \\
-z_3 \t r_{33} & z_0(\t r_{33}+\t r_{11})  & -z_1^* \t r_{11} \\
-z_2^*\t  r_{22} & -z_1 \t r_{11} & z_0(\t r_{11}+\t r_{22})
\end{bmatrix}.
\]
Therein, $z_j=k_1+k_2 e^{iq_j}$ and $z_0=k_1+k_2$. Going further requires introducing a coordinate system and we choose to work in the basis $(\t e_x,\t e_y)$ with $\t e_x = \t r_3$ and $\t e_y$ unitary and directly orthogonal to $\t e_x$. Accordingly, we have
\[
\t r_1 = \begin{bmatrix}
-1/2 \\ \sqrt{3}/2
\end{bmatrix}, \quad
\t r_2 = \begin{bmatrix}
-1/2 \\ -\sqrt{3}/2
\end{bmatrix}, \quad
\t r_3 = \begin{bmatrix}
1 \\ 0
\end{bmatrix},
\]
and
\[
\t r_{11} = \begin{bmatrix}
1/4 & -\sqrt{3}/4 \\ -\sqrt{3}/4 & 3/4
\end{bmatrix},\quad
\t r_{22} = \begin{bmatrix}
1/4 & \sqrt{3}/4 \\ \sqrt{3}/4 & 3/4
\end{bmatrix},\quad
\t r_{33} = \begin{bmatrix}
1 & 0 \\ 0 & 0
\end{bmatrix}.
\]
Also,
\[
q_1 = -q_x/2 +q_y\sqrt{3}/2, \quad
q_2 = -q_x/2 -q_y\sqrt{3}/2, \quad
q_3 = q_x.
\]
\subsection{Dirac cones}
\begin{figure}[ht!]
\centering
\includegraphics[keepaspectratio,width=\textwidth]{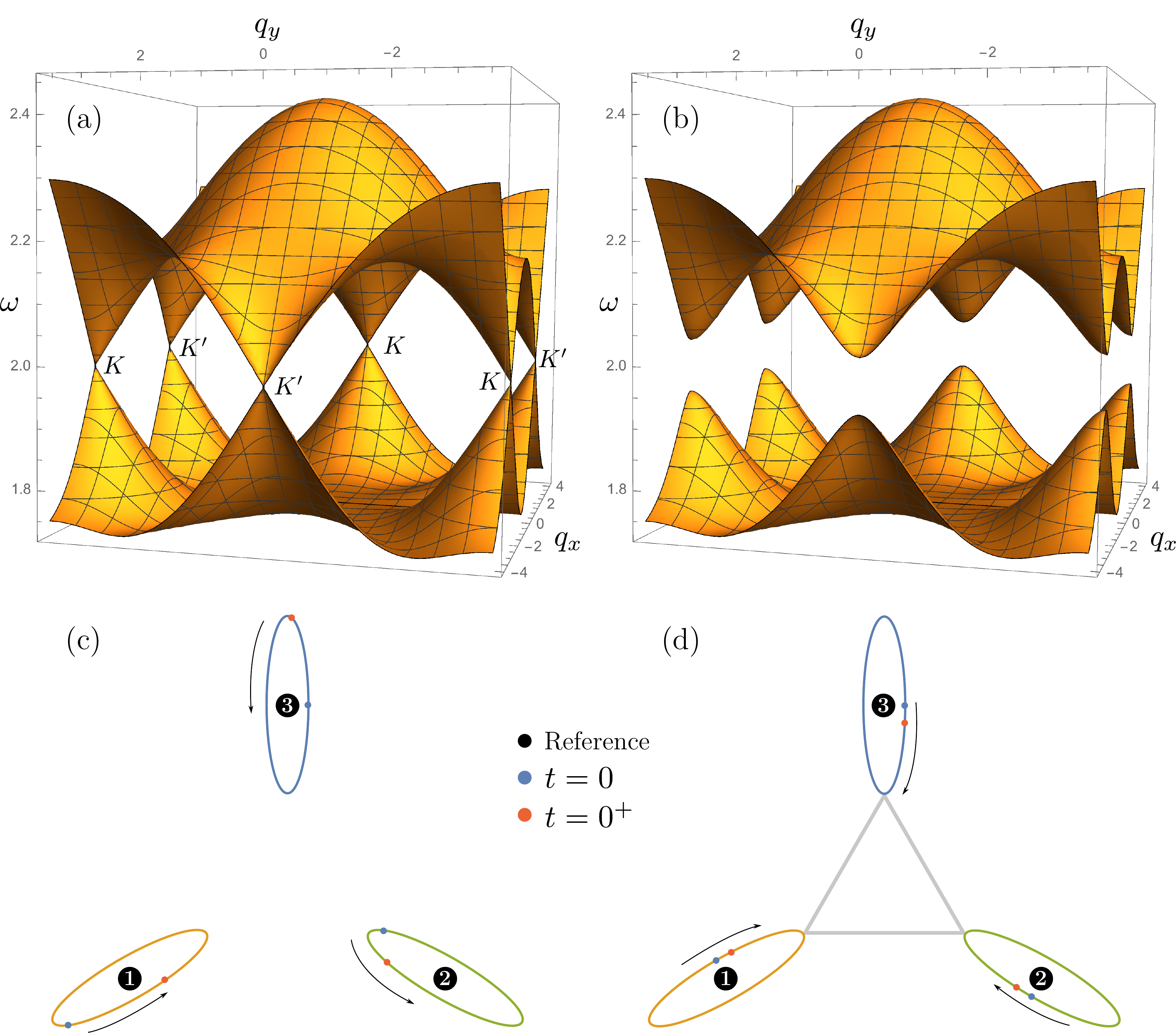}
\caption{Eigenmode analysis: fifth and sixth sheet of the dispersion diagram of a Kagome lattice for (a) $\beta=0$ and (b) $\beta=\pm 0{.}1$; (c,d) orbits of particles within one unit cell respectively corresponding to the eigenmodes $\hat u_K^{\pm}$. Plots valid for $\beta=0$ and asymptotically for $\beta \ll 1$. All along, frequencies are measured in $\sqrt{k/m}$ units.}
\label{fig:kagomeDisp}
\end{figure}
The dispersion diagram of the lattice can be plotted by solving the dispersion relation
\[
\det(\hat H - \omega^2 m I) = 0.
\]
For $\beta=0$, it is seen that the fifth and sixth sheets touch at discrete locations and lead to locally cone-shaped dispersion surfaces known as Dirac cones (Figure~\ref{fig:kagomeDisp}a). The vertices of the cones are located at the corners of the hexagonal Brillouin zone $\t K^\pm_j = \pm\frac{4\pi}{3}\t r_j$. Note that points $\t K^+_j $ are all equivalent and can be deduced from one another by reciprocal lattice translations. Thus, they will be denoted with the same letter $\t K$. Similarly, the remaining $\t K^-_j $ corners are all denoted $\t K'$. All six Dirac cones have the same frequency $\omega_K$ which is further given by
\[
\omega_K^2 = \frac{3(3+\sqrt{5})}{4}\frac{k}{m}.
\]
In other words, at $\t K$ and $\t K'$, $\omega_K^2$ is a double eigenvalue of the dynamical matrix $\hat H$ and has two corresponding eigenmodes respectively called $\hat u_K^{\pm}$ and $\hat u_{K'}^\pm$; see Figure~\ref{fig:kagomeDisp}c,d. All of these modes describe the same elliptical orbits and can be differentiated either by their polarization or by their phase profiles. For instance, at $\t K$, $\Re\left(\hat u_K^{+}e^{-i\omega_K t}\right)$ describes positively oriented elliptical orbits where, within one unit cell, masses $1$, $2$ and $3$ are delayed by a phase of $2\pi/3$ with respect to one another (Figure~\ref{fig:kagomeDisp}c). In comparison, $\Re\left(\hat u_K^{-}e^{-i\omega_K t}\right)$ describes negatively oriented ellipses where the masses within one unit cell are in phase (Figure~\ref{fig:kagomeDisp}d). Note that due to their respective phase profiles, mode $\hat u_K^-$ leads the springs of constant $k_1$ into a state of simultaneous maximum compression (grey triangle on Figure~\ref{fig:kagomeDisp}d) whereas mode $\hat u_K^+$ leads the springs of constant $k_2$ into a similar state (not shown). Similar considerations hold at $\t K'$ and focus will be restricted to $\t K$ points henceforth.

As $\beta$ increases or decreases away from $0$, the upper and bottom parts of the cones separate and a total bandgap opens (Figure~\ref{fig:kagomeDisp}b). It is therefore of interest to investigate whether bandgap opening is accompanied by band inversion phenomena as in the 1D case. But first, an asymptotic model is derived in the vicinity of Dirac cones as it significantly simplifies later derivations.
\subsection{Asymptotic model}
As $\beta$ approaches $0$ and $q$ approaches a Dirac cone, say at a $\t K$ point, eigenmode $\hat u$ approaches the space spanned by the eigenmodes $\hat u^\pm_K$. Thus, to first order, $\hat u$ is given by the expansion
\[
\hat u = s^+ \hat u^+_K + s^-\hat u^-_K + \delta \hat u =
\begin{bmatrix}
\hat u^+_K & \hat u^-_K
\end{bmatrix}
\begin{bmatrix}
s^+ \\ s^-
\end{bmatrix}
+ \delta\hat u
\]
where $s^\pm$ are complex coordinates to be determined and $\delta \hat u$ is a first order correction whereas $\omega$ and $q$ admit the first order expansions
\[
\omega^2 = \omega^2_K + \delta\omega^2, \quad \t q = \t K + \delta \t q
\]
so that $q_j = 4\pi/3 + \delta q_j = -2\pi/3 + \delta q_j$ modulo $2\pi$. Similarly, one has
\[
k_1 = k(1+\beta),\quad
k_2 = k(1-\beta),\quad
z_0 = 2 k,
\]
and
\[
z_j = k(1+e^{i4\pi/3}) +  k\beta(1-e^{i4\pi/3})1+ki\delta q_j e^{i4\pi/3}
\]
thanks to the Taylor series of the exponential function.

The effective motion equation is one that governs the leading order displacements spanned by the coordinates $s^\pm$. It can be obtained by injecting the above expansions into the motion equation~\refeq{eq:MEK} and projecting it onto the subspace spanned by $\hat u_K^\pm$ as in
\begin{multline}
-\begin{bmatrix}
\hat u^+_K & \hat u^-_K
\end{bmatrix}^\dagger
\hat H
\left(
\begin{bmatrix}
\hat u^+_K & \hat u^-_K
\end{bmatrix}
\begin{bmatrix}
s^+ \\ s^-
\end{bmatrix}
+\delta\hat u
\right) \\
= -(\omega^2_K+\delta\omega^2) m 
\begin{bmatrix}
\hat u^+_K & \hat u^-_K
\end{bmatrix}^\dagger
\left(
\begin{bmatrix}
\hat u^+_K & \hat u^-_K
\end{bmatrix}
\begin{bmatrix}
s^+ \\ s^-
\end{bmatrix}
+\delta\hat u
\right).
\end{multline}
Keeping leading order terms yields the result
\[\label{eq:AGE}
\delta\hat H_{K} \begin{bmatrix} s^+\\s^-\end{bmatrix}
=
\delta\omega^2 m \begin{bmatrix} s^+\\s^-\end{bmatrix},
\quad
\delta\hat H_K =
k
\begin{bmatrix}
- a \beta & b (\delta q_x+i\delta q_y) \\
b (\delta q_x-i\delta q_y) & a \beta
\end{bmatrix},
\]
where $a$ and $b$ are the non-dimensional numerical factors
\[
a = 3\frac{5+3\sqrt 5}{20},\quad b = \sqrt 3\frac{5+\sqrt 5}{20},
\]
and $(\delta q_x,\delta q_y)$ are the coordinates of the correction $\delta \t q$ in the basis $(\t e_x, \t e_y)$. Note that the calculation of the eigenmodes $\hat u^\pm_K$ is straightforward to carry using a numerical routine or a symbolic computation software.
\subsection{Band inversion}
\begin{figure}[ht!]
\centering
\includegraphics[keepaspectratio,width=\textwidth]{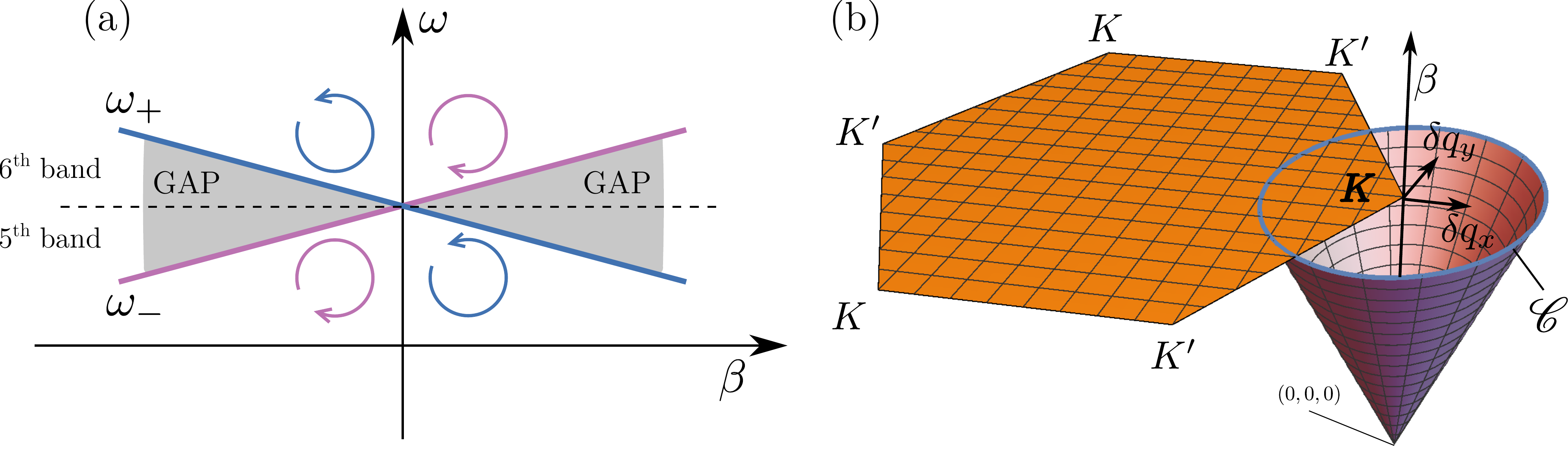}
\caption{Band inversion at $\t q=\t K$: (a) bandgap frequencies as a function of $\beta\ll 1$: as $\beta$ changes sign the eigenmode of the fifth/sixth band reverses its polarization; (b) the 3D space of parameters $(\delta q_x,\delta q_y,\beta)$: the cone's vertex is at the origin and its height is equal to $\beta$. The geometric phase $\Delta$ accumulated along $\curve$ is minus half of the solid angle of the cone and rapidly converges to $\pm \pi$ for $\beta\ll 1$.} 
\label{fig:solidAngle}
\end{figure}

The dispersion relation in the vicinity of $\t K$ is
\[
\frac{m^2}{k^2}(\omega^2-\omega_K^2)^2 - a^2\beta^2 = b^2(\delta q_x^2 + \delta q_y^2)
\]
and results from the condition of zero determinant applied to $\delta\hat H$ rather than~$\hat H$. For $\beta=0$, the above equation indeed describes a cone (Figure~\ref{fig:kagomeDisp}a). For $\beta\neq 0$, the cone separates into two disconnected hyperbolic sheets and a bandgap opens between the frequencies
\[
\omega^2_{\pm} = \omega^2_K\mp\frac{k}{m}a\beta
\]
where $\omega^2_{\pm}$ are the corrected eigenfrequencies of the modes $\hat u_K^\pm$ (Figure~\ref{fig:kagomeDisp}b). Thus for $\beta<0$, mode $\hat u_K^-$ has a lower frequency than $\hat u_K^+$ and they belong to the fifth and sixth dispersion sheets respectively. As $\beta$ increases, $\omega_-$ increases whereas $\omega_+$ decreases until they meet at $\omega_K$ for $\beta=0$ where the sheets touch and the gap closes. Beyond that, for $\beta >0$ it is $\hat u_K^-$ that has the higher frequency and belongs to the sixth sheet whereas $\hat u_K^+$ now has the lower frequency and belongs to the fifth sheet; see Figure~\ref{fig:solidAngle}a. The described band inversion phenomenon is qualitatively identical to the one that takes place in the 1D case. As a matter of fact, for $\beta > 0$, i.e., $k_1>k_2$, mode $\hat u_K^-$ leads the stiffer $k_1$ springs into a state of maximum compression due to its in-phase profile (see gray triangle on Figure~\ref{fig:kagomeDisp}d) and is therefore more energetic, i.e., of a higher frequency, than mode $\hat u_K^+$. Compared to the 1D case, the novelty is in the accompanying polarization that emerges in a 2D setting.

Furthermore, band inversion can here too be characterized by a geometric phase
\[
\Delta = \Im\int_\curve \scalar{\hat u_K^+,\t\partial_{\delta q} \hat u_K^+}\cdot \dd \delta \t q
\]
understood as a path integral along a loop, say a circle, $\curve$ centered on $\t K$ in $\t q$-space. Also known as a Berry's phase, $\Delta$ can be obtained using Stokes theorem as half of the solid angle of a surface $S$ subtended by $\curve$ as seen from point $(0,0,0)$ in the $(\delta q_x, \delta q_y, \beta)$-space \citep{Berry1984}; see Figure~\ref{fig:solidAngle}b. Given that a plane that does not contain the origin has a solid angle of $\pm 2\pi$ and taking $\beta$ to be sufficiently small, it comes that the winding number $\nu=\Delta/\pi$ is quantized and can only take two values
\[
\nu = +1 \quad\text{for}\quad \beta > 0,\quad
\nu = -1 \quad\text{for}\quad \beta < 0.
\]
By the bulk-edge correspondence principle, two gapped Kagome lattices with different winding numbers, i.e., with opposite $\beta$ are therefore expected to host at the interface localized eigenmodes whose frequencies fall inside the common bandgap. These topological Stoneley waves are investigated next.
\subsection{Topological Stoneley waves: smooth interface}
\begin{figure}[ht!]
\centering
\includegraphics[keepaspectratio,width=\textwidth]{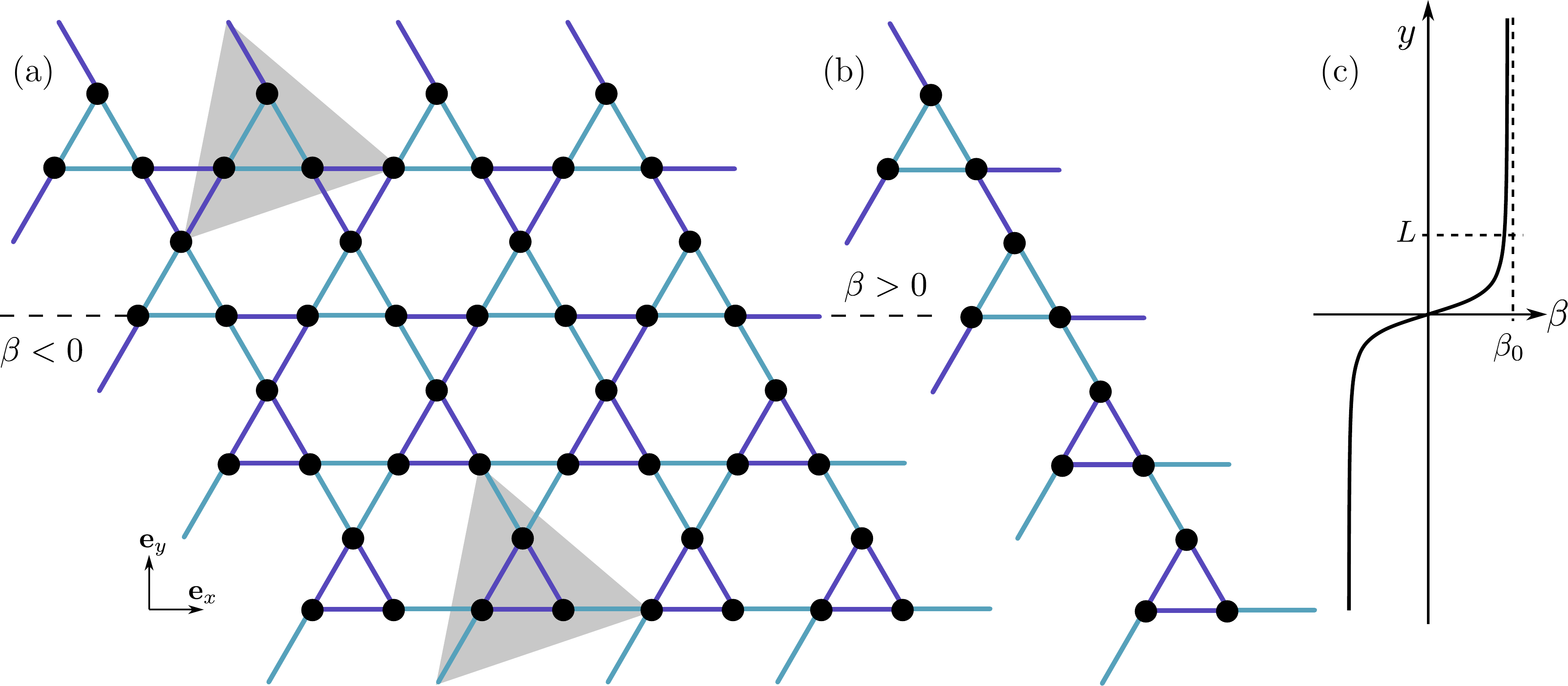}
\caption{(a) An interface ($y=0$, dashed line) separates two topologically distinct Kagome lattices: two unit cells (shaded triangles) below and above the interface have opposite contrasts $\beta$; the system is periodic in the $x$-direction. (b) A super cell summarizing the periodic geometry: left and right boundaries are subjected to Floquet-Bloch conditions with wavenumber $q=K+\delta q_x e_x$; top and bottom boundary conditions are irrelevant for modes localized at the interface in the limit where the sample has enough unit cells in the $y$-direction (around $5b/(a\beta_0)$). (c) The contrast profile $\beta$: parameter $L$ quantifies the thickness of the interface.}
\label{fig:interface}
\end{figure}

An interface located at $y=0$ between two topologically distinct Kagome lattices occupying the two half-planes $y>0$ and $y<0$ can be described by a profile $\beta=\beta(y)$ such that $\beta(y<0)$ and $\beta(y>0)$ are non-zero and have constant and opposite signs, say $\beta(y<0)<0$ and $\beta(y>0)>0$. In addition, we shall assume that $\beta$ remains small and varies slowly in compliance with the prerequisites of the derived homogenized model. Other cases will be investigated in a later section. Let then, for the sake of example,
\[
\beta(y) = \frac{y/L}{\sqrt{1+y^2/L^2}}\beta_0
\]
be the contrast profile, $L$ being a distance characterizing the width of the interface between the two topologically distinct lattices and $\beta_0>0$ being maximum contrast attained at infinity. Figure~\ref{fig:interface} illustrates the adopted configuration.

Writing the homogenized motion equations~\refeq{eq:AGE} in differential form (simply map $i\delta q_y \mapsto \partial_y$),
\[
\begin{split}
- a \beta(y)s^+ + b (\delta q_x+\partial_y)s^- &= \frac{m\delta\omega^2}{k} s^+,\\
a \beta(y)s^- + b (\delta q_x-\partial_y)s^+ &= \frac{m\delta\omega^2}{k} s^-,
\end{split}
\]
solutions can be looked for in the form
\[
s^\pm(y)
=
S^\pm \exp\left(-\frac{a}{b}\int_0^y\beta(y)\dd y\right).
\]
Note that due to the fact that $\beta\to \pm\beta_0$ at $\pm\infty$, it is guaranteed that $s^\pm$ will decay exponentially away from the interface. Substituting back into the motion equations, it is deduced that the waves amplitudes satisfy $S^+ = -S^- \equiv S$ and that the wavenumber is solution to the Stoneley wave dispersion relation
\[
\delta q_x = -\frac{m\delta \omega^2}{bk} = -\frac{2m\omega_K}{bk}\delta\omega.
\]
\begin{figure}[ht!]
\centering
\includegraphics[keepaspectratio,width=\textwidth]{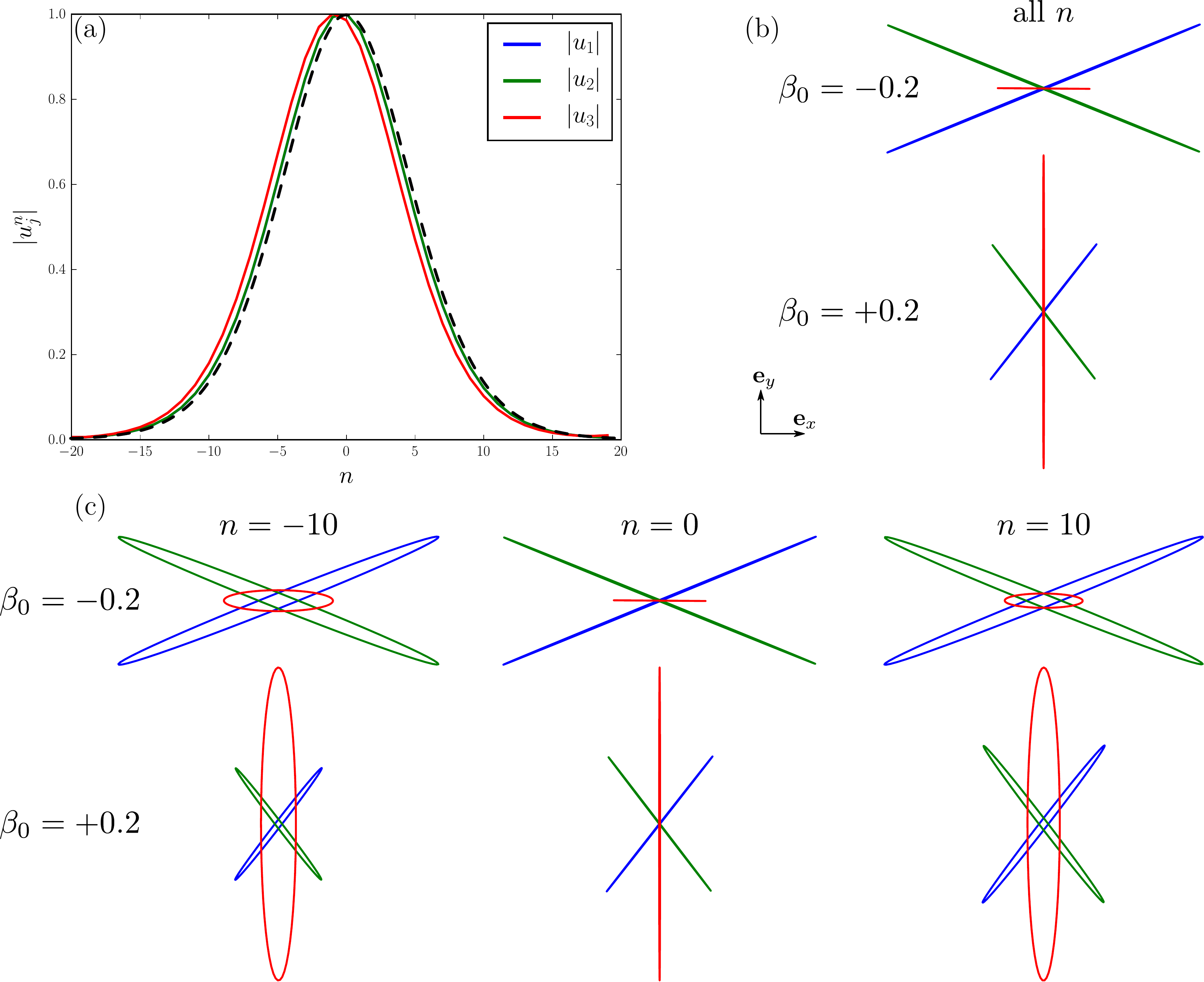}
\caption{Topological Stoneley waves at $(\t K,\omega_K)$ along a smooth interface. (a) Spatial profile: Normalized displacements amplitude calculated numerically (three solid lines) decay exponentially in agreement with the asymptotic model (one dashed line). (b) Asymptotic normalized trajectories of the masses within one unit cell. (c) Simulated trajectories. The asymptotic model is most accurate near the interface ($n=0$) where $\beta$ is the smallest.}
\label{fig:interfaceModes}
\end{figure}

In conclusion, and more generally, the total displacement field of the Stoneley wave is given by
\[
\hat u(x,y,t) = S \exp\left(\mp\frac{a}{b}\int_0^y\beta(y)\dd y\right) \left(\hat u^+_K \mp \hat u^-_K\right)e^{i(4\pi x/3-\omega_K t)}e^{i(\delta q_x x - \delta \omega t)}
\]
and its dispersion relation is
\[
\delta q_x = \mp\frac{2m\omega_K}{bk}\delta\omega.
\]
Therein, the first (resp. second) sign corresponds to cases where $\beta_0$ is positive (resp. negative).

\begin{figure}[ht!]
\centering
\includegraphics[keepaspectratio,width=\textwidth]{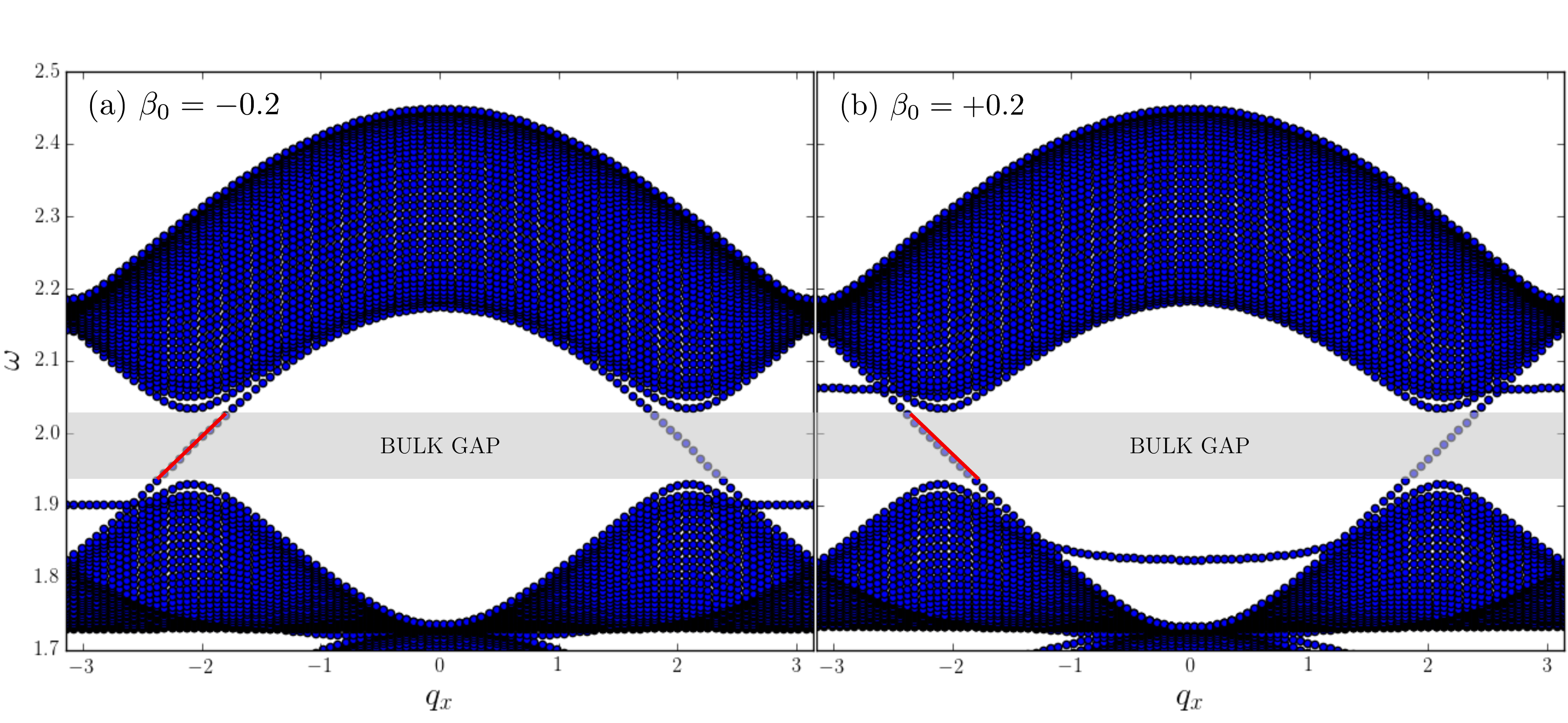}
\caption{Dispersion relation of Stoneley waves along a smooth interface obtained by numerical eigenmode analysis (blue dots) and by asymptotic analysis (red line) for (a) negative and (b) positive $\beta_0$.}
\label{fig:StoneleyDisp}
\end{figure}

Eigenmode analysis conducted numerically confirm our analytical predictions. For $\beta_0=\pm 0{.}2$, the decay speed of Stoneley waves is around $a\beta_0/b = 0{.}56$ meaning that a sample of more than $20$ unit cells in the $y$-direction is practically infinite for our purposes. Correspondingly, in that limit, top and bottom boundary conditions are irrelevant. As for the interface thickness $L$, $\t q$ being in the vicinity of $\t K$ with $q_x\approx -2\pi/3$ corresponding to modes periodic across three unit cells, the interface should count no less than three unit cells for the homogenized model to apply, that is $L\geq 3\sqrt{3}/2$. The simulations were carried on a sample of $40$ unit cells in total in the $y$-direction and with $L=10\sqrt{3}/2$, under fixed boundary conditions. Figure~\ref{fig:interfaceModes} depicts the spatial profile as well as mass trajectories of the predicted topological Stoneley waves whereas dispersion diagrams are plotted in Figure~\ref{fig:StoneleyDisp}. Overall, satisfactory agreement between the numerical and asymptotic models is observed. Note that by time reversal symmetry, the same results hold at $\t K'$ but are not illustrated here.

As we move away from the interface, the asymptotic model no longer predicts accurately the elliptical trajectories of the masses and only accounts for their major axes (Figure~\ref{fig:interfaceModes}b,c). Indeed, in the limit $\beta \to 0$, $\hat u^+_K$ and $\hat u^-_K$ describe the exact same elliptical trajectories but as $\beta$ increases, the trajectories deform in different manners and their addition/subtraction no longer produce the linear profiles of Figure~\ref{fig:interfaceModes}b but the elliptical ones of Figure~\ref{fig:interfaceModes}c. Taking these effects into account is possible by recalculating and correcting the expressions of $\hat u^\pm_K$ for finite non-zero contrast $\beta$ but will not be pursued here.

Three key features distinguish the present topological Stoneley waves from their classical predecessors. First, their frequencies traverse a total bulk bandgap making their scattering into the bulk impossible (Figure~\ref{fig:StoneleyDisp}). Second, their decay speed within the gap is frequency independent as it only depends on the geometry of the lattice through the ratio $a/b$ and on an averaged value of the contrast $\int\beta(y)\dd y$. Third and last, their hosting interface is polarized: exchanging the top and bottom half-planes, or equivalently applying $y\mapsto -y$ drastically changes the orbits wherein mass $3$ oscillates parallel to the interface in one case and orthogonal to it in the other case (Figure~\ref{fig:interfaceModes}b,c).
\subsection{Topological Stoneley waves: discontinuous interface}
It can be argued that the Stoneley waves characterized in the previous subsection can be classically explained as waves guided within a thin conducting layer $\abs{y}\ll L$ with $\beta=0$ and where the bulk bandgap vanishes surrounded by two isolating half-planes $\abs{y} > L$ with $\beta\neq 0$ without recurring to the above topological and asymptotic tools. The purpose of this second example is to show that this is not the case. That is, even in the absence of layers where $\beta$ vanishes, gapless Stoneley waves will exist along interfaces separating topologically distinct lattices. Thus, let the profile
\[
\beta(y>0) = \beta_0 \neq 0, \quad \beta(y<0) = -\beta_0
\]
substitute the previously smooth contrast profile. The geometric description of Figure~\ref{fig:interface}a,b is maintained.
\begin{figure}[ht!]
\centering
\includegraphics[keepaspectratio,width=\textwidth]{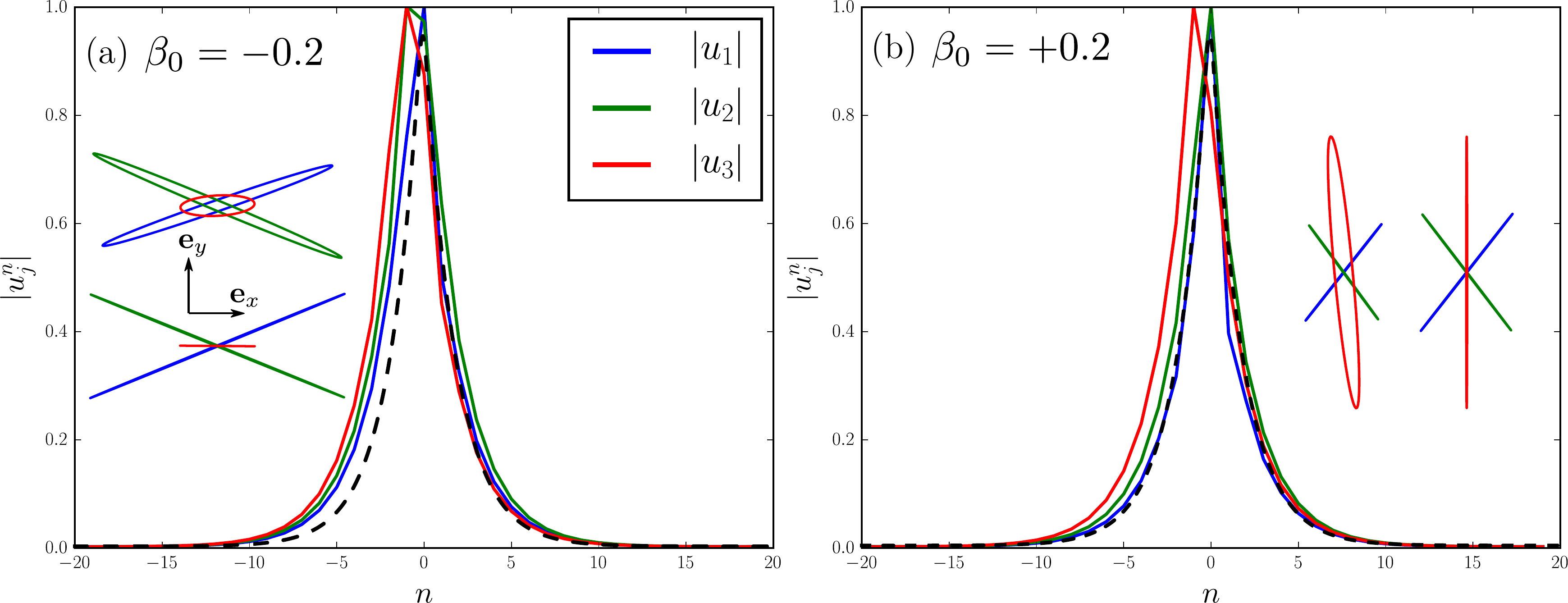}
\caption{Topological Stoneley waves along a discontinuous interface at $(K,\omega_K)$: Normalized displacements amplitude calculated numerically (three solid lines) decay exponentially in agreement with the asymptotic model (one dashed line). Insets show the simulated (top and left) and asymptotic (bottom and right) trajectories of the masses within one unit cell at the interface ($n=0$).}
\label{fig:interfaceModesDisc}
\end{figure}

\begin{figure}[ht!]
\centering
\includegraphics[keepaspectratio,width=\textwidth]{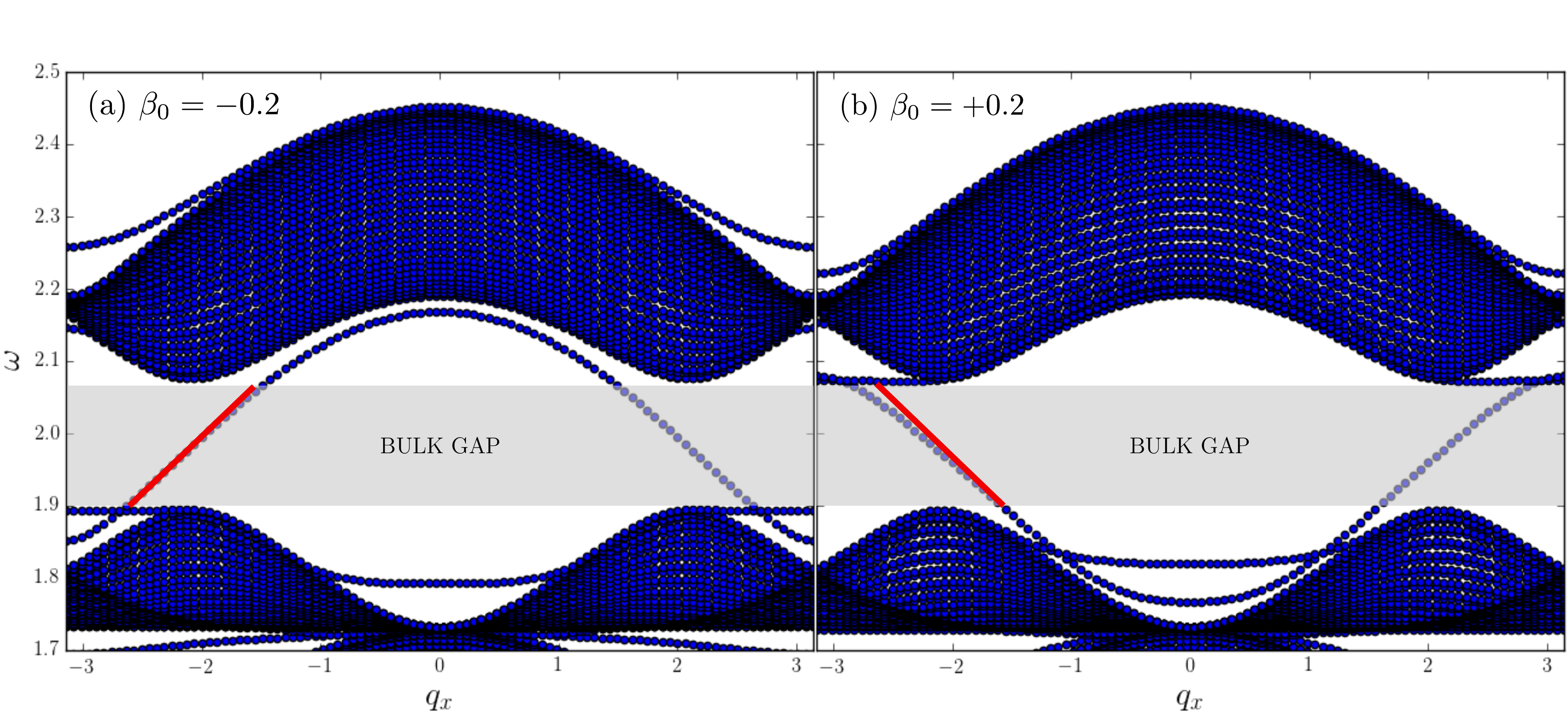}
\caption{Dispersion relation of Stoneley waves along a discontinuous interface obtained by numerical eigenmode analysis (blue dots) and by asymptotic analysis (red line) for (a) negative and (b) positive $\beta_0$.}
\label{fig:StoneleyDispDisc}
\end{figure}

Then, the total displacement field of the Stoneley wave is similarly given by
\[
\hat u(x,y,t) = S \exp\left(\mp\frac{a}{b}\beta_0\abs{y}\right) \left(\hat u^+_K \mp \hat u^-_K\right)e^{i(4\pi x/3-\omega_K t)}e^{i(\delta q_x x - \delta \omega t)}
\]
with the same sign convention and the same dispersion relation as before. Mode shapes and dispersion diagrams are plotted on Figures~\ref{fig:interfaceModesDisc} and~\ref{fig:StoneleyDispDisc} confirming the existence of gapless Stoneley waves localized at the interface.

In conclusion, although all layers $y$ are insulating and $\beta$ never vanishes, the gap still closes in a sense at $y=0$ in order to accommodate the band inversion phenomenon that occurs there and that was characterized topologically earlier, thus allowing for the emergence of Stoneley waves.
\subsection{Backscattering at corners}
Time reversal symmetry ensures that each time there is a Stoneley wave in the vicinity of point $\t K$, there is a Stoneley wave, with opposite polarization and opposite group velocity, in the vicinity of point $\t K'$. It is nonetheless remarkable that it is possible to deal with these waves separately as they are uncoupled. This is most apparent on the dispersion diagrams of either Figure~\ref{fig:StoneleyDisp} or~\ref{fig:StoneleyDispDisc}. Therein, it is seen that point $\t K$, located at $q_x=-2\pi/3$, and point $\t K'$, located at $q_x=2\pi/3$, are separated by a wavenumber of $4\pi/3$ which falls significantly shorter than the structural wavenumber equal to $2\pi$. Accordingly, Bragg reflection or scattering coupling points $\t K$ and $\t K'$ is negligible.
\begin{figure}[ht!]
\centering
\includegraphics[keepaspectratio,width=\textwidth]{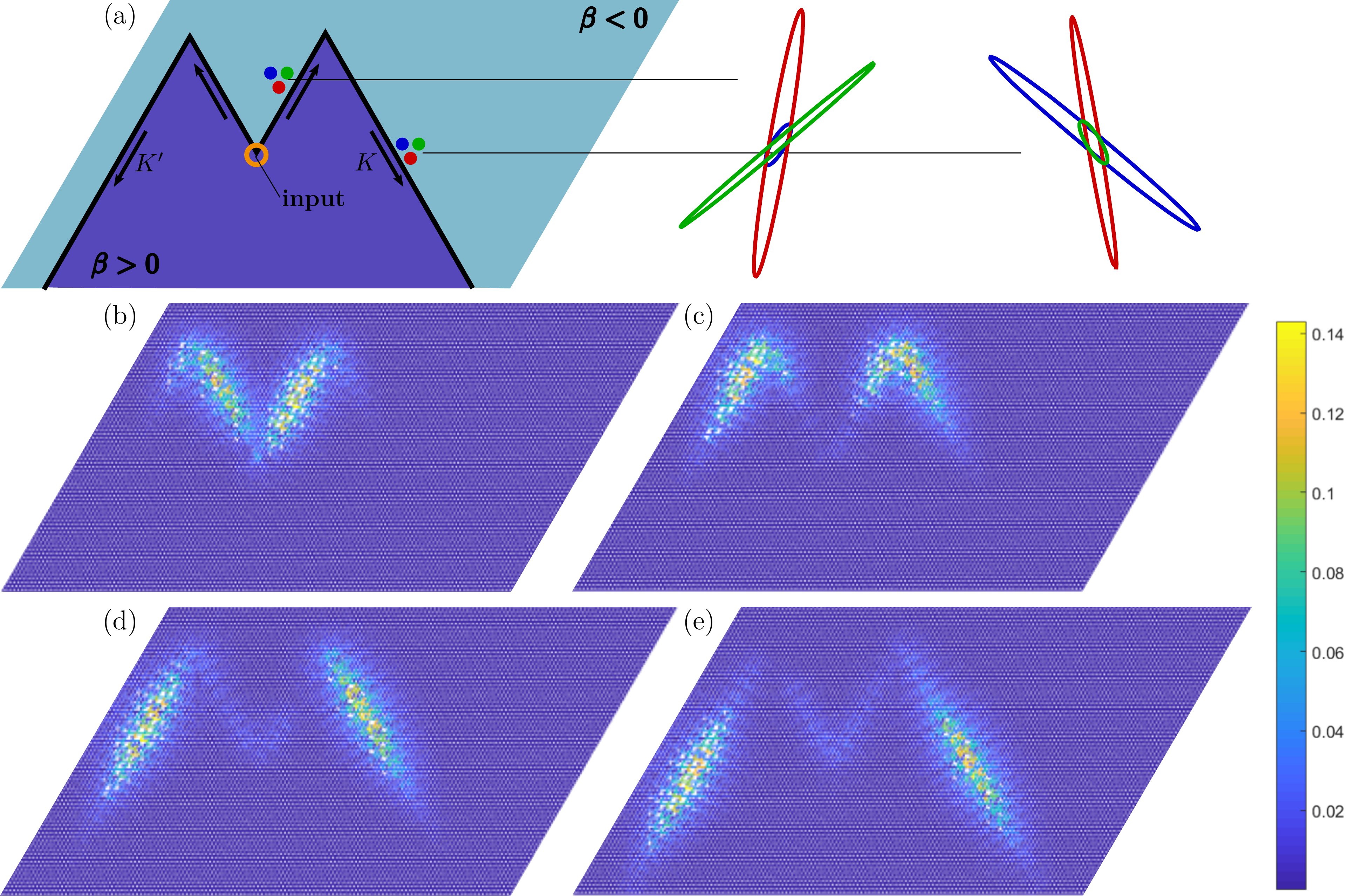}
\caption{Absence of backscattering at corners. (a) Geometry: an M-shaped interface separating two topologically distinct Kagome lattices; an input signal is transmitted with no backscattering at corners; waves going right (resp. left) belong to point $K$ (resp. $K'$); simulated orbits are depicted for two pinned unit cells. (b-e) Transient numerical simulations: snapshot profile of normalized velocity amplitude (color bar) respectively at $t=150, 200, 250$ and $300$. Numerical simulations are carried over the geometry described in (a) under free boundary conditions; the contrast parameter is $\beta=\pm0{.}2$ and the input is a 60-cycle tone-burst body force centered on $\omega = 1{.}95$; the orbits in (a) are described between $t-\pi/\omega$ and $t+\pi/\omega$ as the wave passes by the corresponding unit cell at $t=150$ (b) and  $t=250$ (e).}
\label{fig:corners}
\end{figure}

The argued uncoupling has a remarkable consequence in terms of absence of backscattering of topological Stoneley waves at corners. Consider the M-shaped interface depicted on Figure~\ref{fig:corners}a. A loading applied at the input position will emit two waves. As $\beta$ is positive below the interface and negative above it, the wave going right belongs to point $\t K$ and the one going left belongs to point $\t K'$; see Figure~\ref{fig:StoneleyDispDisc}b. The corner, featuring interfaces of type $\t K$/$\t K'$, will not couple the $\t K$ and $\t K'$ Stoneley waves that will therefore follow the abrupt change in the interface with negligible backscattering (Figure~\ref{fig:StoneleyDispDisc}b-e). The simulation was carried in time domain over a sample of $60\times 40$ unit cells under free boundary conditions. The selected contrast parameter is $\beta=\pm0{.}2$ and is constant with an abrupt change at the M-shaped interface. The loading is a narrow-band 60-cycle tone-burst horizontal body force centered on $\omega = 1{.}95$ and applied at the center tip of the interface as illustrated.

In contrast, an interface parallel to $\t r_1 - \t r_2$ will mix the $\t K$ and $\t K'$ Stoneley waves as $\scalar{\t r_1-\t r_2,\t K-\t K'}=0$; that is, the $\t K$ and $\t K'$ wavenumbers projected onto the interface parallel to $\t r_1-\t r_2$ are in fact identical. Consequently, said interface will couple the $\t K$ and $\t K'$, i.e., left and right, Stoneley waves and significant backscattering is to be expected at defects, corners or otherwise, in that case.
\section{Conclusion}
The behavior of inhomogeneous Kagome lattices in the vicinity of Dirac cones can be classified based on a topological invariant quantifying a geometric phase and representing the change in the phase profile of a pair of elliptically polarized stationary eigenmodes with opposite handedness. Two classes of topologically distinct Kagome lattices thus emerge and are guaranteed, whenever they share an interface, to host a couple of gapless Stoneley waves in their common bulk bandgap. This constitutes the first adaptation of the so-called ``quantum valley Hall effect'' to in-plane elasticity.

It is important to highlight that the bulk-edge correspondence principle exemplified in this manner is not absolute. For instance, in the 1D scenario, the quantization of the winding number is a direct result to the hypothesis of equal masses $m_1=m_2$. In the 2D scenario, the geometric phase was proven topologically invariant, i.e., quantized, asymptotically in the limit $\beta\ll 1$. Further, the inhomogeneous Kagome lattice remained $C_3$ symmetric eventhough $C_6$ symmetry was lost. Correspondingly, the existence and robustness of Stoneley waves are not absolute either and remain subject to these symmetry conditions. Therefore, future efforts quantifying the extent of this symmetry protection against uncertainty and defects remain much needed.
\section*{Acknowledgement}
This work is supported by the Air Force Office of Scientific Research under Grant No. AF 9550-15-1-0061 with Program Manager Dr. Byung-Lip (Les) Lee, the Army Research office under Grant No. W911NF-18-1-0031 with Program Manager Dr. David M. Stepp and the NSF EFRI under award No. 1641078.

\end{document}